# A Markovian model of evolving world input-output network

Vahid Moosavi[1*], Giulio Isacchini[2]

[1] Chair for Computer Aided Architectural Design, Department of Architecture, ETH Zurich, Switzerland

[2] Department of Physics, ETH Zurich, Switzerland

[*] Corresponding author
E-mail: svm@arch.ethz.ch (VM)

# Abstract

The initial theoretical connections between Leontief input-output models and Markov chains were established back in 1950s. However, considering the wide variety of mathematical properties of Markov chains, so far there has not been a full investigation of evolving world economic networks with Markov chain formalism. In this work, using the recently available world input-output database, we investigated the evolution of the world economic network from 1995 to 2011 through analysis of a time series of finite Markov chains. We assessed different aspects of this evolving system via different known properties of the Markov chains such as mixing time, Kemeny constant, steady state probabilities and perturbation analysis of the transition matrices.

First, we showed how the time series of mixing times and Kemeny constants could be used as an aggregate *index of globalization*.

Next, we focused on the steady state probabilities as a measure of *structural power of the economies* that are comparable to GDP shares of economies as the traditional index of economies welfare.

Further, we introduced two measures of systemic risk, called *systemic influence* and *systemic fragility*, where the former is the ratio of number of influenced nodes to the total number of nodes, caused by a shock in the activity of a node and the latter is based on the number of times a specific economic node is affected by a shock in the activity of any of the other nodes.

Finally, focusing on Kemeny constant as a global indicator of monetary flow across the network, we showed that there is a paradoxical effect of a change in activity levels of economic nodes on the overall flow of the world economic network. While the economic slowdown of the majority of nodes with high structural power results to a slower average monetary flow over the network, there are some nodes, where their slowdowns improve the overall quality of the network in terms of connectivity and the average flow of the money.

## Introduction

The mathematical beauty of computational algebraic methods such as Markov chains is that they are domain free. This means that having a proper size of observed data and enough computational power they fit very well into many application domains, while unlike many domain specific models, they do not ask for domain specific prior-knowledge. For example, they assume that the rules of interactions among agents (being economic agents or drivers in a transportation network or words in a spoken language), are embedded in the traces of their real interactions, while in traditional rule based or agent based simulations, one needs to specify features and the rules of interactions among those agents beforehand. On the other hand, algebraic methods are data demanding and because of this, Markov chains for example that were introduced in 1906 [1], did not get that much of attention before the advent of computers in 1950s and finally in the late 1990s, Markov chains were applied in large scale problems such as in PageRank algorithm in Google search engine [2]. In principle, the same argument holds for the recently successful field of "*representation learning*" or the so-called "*deep learning*", where having large amount of data set along with a series of stacked algebraic operators one can come up with highly sophisticated hierarchical representations of complex phenomena [3].

In this work our focus is on Markov chains and their applications on evolving economic networks. A Markov chain is a data driven formalism to its underlying dynamical system, where we only need some real observations and usually no prior rules of interactions among the agents or the states of that system. Nevertheless, with this formalism one can benefit from the many interesting mathematical properties of Markov chains such as their steady state probability distribution [4], Kemeny constant [5], recurrence time and mixing time [6], mean first passage times [7] and the sensitivity analysis of the underlying networks through perturbation of the transition matrix [8-10]. Of course, one should be very careful with the prior assumptions in a Markov chain such as its structuralist view to the problem, the issues of memory, the linearity of the operator, the assumptions about closed-ness of the state space in discrete chains, etc.

In the domain of economic and financial applications, especially after the financial crisis of 2008, the notions of *networked economy*, *complexity and systemic risk* are gaining increasing importance [11-16]. Comparing to classical economic models, which are mainly based on the assumptions of independent agents, network based economics is focused on the interaction between agents.

Nevertheless, networks are not new topics in economics. For example, one can refer to the works of Leontief on the so-called, input-output tables [17] within 1940s, for which he won a Nobel Prize in economics. An input-output table in fact is a network, where nodes are the segments of an economy (i.e. different industries within a country) and the edges are the monetary flows of goods within these nodes. Input-output tables can be seen as a system of equations where the solution (if exists) is considered as the equilibrium price of products in order to keep the economic network stable.

Related to the our work, Solow in 1952 [18] discussed the connections between Leontief input-output models and Markov chain formalism, where he investigated the required conditions for finding a stable solution (i.e. balanced prices) for the underlying system of equations. Further, the authors in [19] modeled input-output models as absorbing Markov chains based on either the flow of materials or the flow of money.

In this work, based on the recently available data set, called World-Input-Output-Database (WIOD) [20], we investigate several other properties of Markov chains on a time varying global economic network.

In the next section, we briefly describe the data set we used in this work. Next, we describe the proposed Markovian model and those properties we applied to analyze the global economic network. Finally, we show the results and discuss the potential future directions.

# Materials and methods

## World input output network

The World-Input-Output-Database (WIOD) represents a network of two types of nodes. The first type of node, $I$, corresponds to a specific industrial sector within an economy. Each industry, based on some inputs from other industries, produces some products and sells them to other intermediate industries and final consumers that are call households and the governments. These households, together with the government of each economy, represent an additional kind of node, $G$. This node participates in the money flow through the network by consumption of final products, and by receiving money consisting of taxes and value added coming from the corresponding industries working in that economy. This process can be visualized in a weighted digraph structure, where an industrial sector $j$ of a specific economy $i$ is defined as $E_i I_j$. Further, we assign one node for the governments and households within each economy that from now on we refer to by $E_i G$. In this manner, each input output table is represented as a closed system, which makes it suitable for Markov Chain formalism.

Fig 1 shows a schematic view of a closed network with two economies, each with one unique industry and one node representing the government and the households together. As it is shown in this figure, in WIOD data set, due to aggregation of flows within industries, there are explicit self-loops for industry nodes. Further, from now on we assume that the edges are representing the flow of money. As we will show later, it is also possible to easily aggregate the flows over each economy in order to come up with the measures at the level of economies.

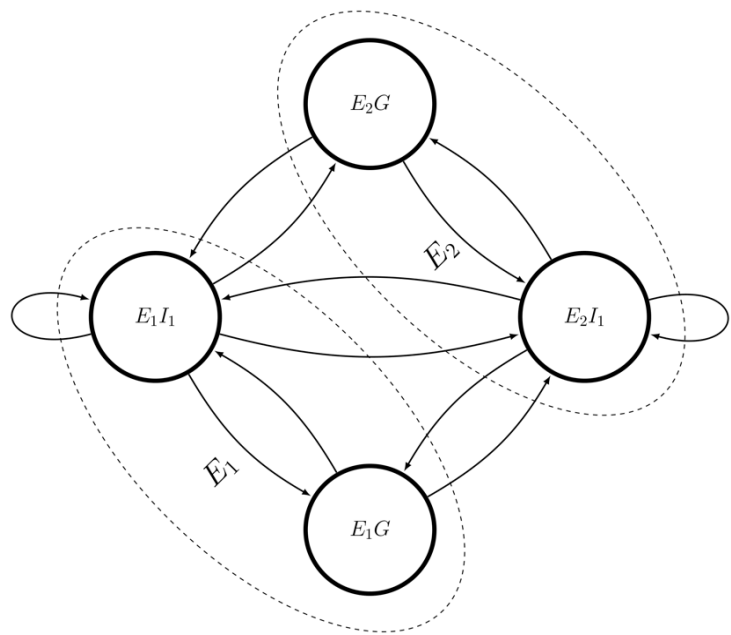

**Fig 1. A schematic view of a closed economic network**

There are two economies and one industry within each economy, and one node for the government and households within each economy. The edges represent the flow of money between nodes.

In the WIOD that we used in this study, there are 35 industries within 41 economies (27 EU countries and 13 major economies in other regions) plus the rest of the world (RoW) as one economy. A complete list of industries and economies can be found in [20]. Considering the 35 industries within each economy plus one node for each government and households (together in one node), there are 1476 nodes for each year. While the flows (i.e. the edges and their values) change from year to year, the same structure repeats for 17 years from 1995 to 2011, which makes it suitable for trend analysis. WIOD is a valuable data set that has been used in several recent studies, including identification of global value chains and trade fragmentation [21,22] and global environmental accounting in ecology and resources management [23]. From a network analytics point of view recently there has been a work on this data set, where several network based measures such as different centrality measures and clustering measures of the world economic network have been studied [24]. In this work we applied several properties of Markov chains on this time varying network.

# The proposed Markovian model of the world economic network

As mentioned before, the formalism of Input Output (IO) models by Markov chain has an old history back to 1952 [18] and more recently to [19] who modeled an open IO models as absorbing Markov chains. In this work, a closed IO network is studied, which can be translated naturally to a regular Markov chain with no absorbing states.

More explicitly, an IO table of a specific year in the WIOD is an asymmetric non-negative squared matrix $W$ whose elements $w_{ij}$ correspond to the flow of money from a node $j$ to a node $i$. A stochastic matrix $T$ can be directly associated to $W$ by column normalization, thus a specific element $t_{ij}$ of $T$ is defined by:

$$t_{ij} = \frac{w_{ij}}{\sum_i w_{ij}} \qquad (1)$$

The elements $t_{ij}$ can be interpreted either as the relative flow of money between nodes or as the probability for a random walker to move from one node to another. Since there is only one table per year, these probabilities are the annual average values. As a result, for each year, we assume a single discrete time-homogeneous Markov chain model with a corresponding stochastic matrix $T_y$ with $y \in \{1995,\ldots,2011\}$.

Our interest in this work is mainly on the dynamics of the world economic network over time, where the $T_y$ matrices are changing for each year. Therefore, we are facing a time inhomogeneous Markov chain. However for every year $y$ the stochastic matrix is well defined and its properties can be used to characterize the global economic network and follow its evolution through 17 years. In particular three specific properties have been chosen:

**Steady State Vector:** is the first eigenvector of $T$ defined by:

$$\pi = T\pi \qquad (2)$$

One can easily estimate this vector using power iteration method, starting with any initial random vector. This vector is sometimes called *Eigenvector centrality,* however this terminology has not been used in this work. In our case, $\pi$ is a normalized one-dimensional vector with the same size as the number of nodes in the global economic network (i.e. 1476 nodes). Its values can be interpreted as the expected long-term relative amount of money within each government or each industry. As we will see later, while it is common in network studies to use centrality measures for ranking of the nodes (e.g. PageRank algorithm [2]), here these values are highly comparable to annual GDP shares of the economic nodes. It is important to note that since the underlying dynamical system in our global economic network cannot be explained with one fixed transition matrix, one cannot claim that the global economic network reaches the steady state within the scope of one year. But at the same time, taking stationary probabilities as a kind of *structural property* of each node, the comparisons of their values over time reveals interesting features of this evolving global economic network.

**Mixing time**: It can be measured as the average number of steps that a Markov chain takes from any random initial state in order to reach its steady state [6]. Mixing time is a very good global measure, which shows how connected the network is. In principle, if a chain has more local loops or disconnected regions that is difficult to enter or leave, mixing time will be longer. In the context of global economic network this can be considered as an index of globalization, where higher values of mixing times shows less connected network and vice versa. In this work, the mixing time of each year's transition matrix is calculated through the average number of iterations in the power iteration method.

**Kemeny constant:** Similar to mixing time, this is another global measure of the Markov chains, which shows the average expected time from any given state (node) to a random state (node). Interestingly, this value is constant over different states of a given Markov chain and therefore it can be considered as an intrinsic feature of a chain. Similar to mixing time, this constant can be a good indicator of the connectivity of the underlying network. Therefore, as a hypothesis we expect that corresponding

Kemeny constants of different Markov chains for different years should form a decreasing pattern over time, which indicates a faster flow of money and more development of the global economic network within the years 1995 to 2011. Along the same line, there is another interesting property of Markov chains, called Mean First Passage Time (MFPT) [7], which indicates the expected time for a Markov chain to transit from specific node to another specific node. In the context of economic networks, this measure can be used to analyze the inter-relationships between two specific industries within or across a value chain. However, we did not use this measure in this work.

Calculation of Kemeny constant of a each Markov chain is very straight forward. As shown in [5], the eigenvalues $\lambda_2, \ldots, \lambda_n$ of $T$ other than 1 can be used to compute the Kemeny constant as follows:

$$K(T) = 1 + \sum_{i=2}^{n} \frac{1}{1-\lambda_i} \qquad (3)$$

**Sensitivity analysis of transition matrices**

By perturbing the values of the transition matrices, one can analyze the effect of each node on the other nodes. There are many different approaches for perturbation analysis of Markov chains within the literature such as [8-10]. A common way for perturbation analysis is to change the transition probabilities by small random noises, while the sum of these noises is equal to zero. In this way the transition matrix will remain stochastic.

However in this work, since we are ultimately interested in defining risk measures attributed to individual economic nodes, we choose a different procedure of perturbation, repeated for all the nodes. As described in [10], we analyze the effect of slowing down the activity level of one specific economic node on all the other nodes by the following procedure.

If we want to change the activity of one node by $\alpha$ percent we multiply all the outflow and inflow rates of that node by $1 + \alpha/100$ and then we normalize all the affected columns. After this change, we have a new transition matrix.

One interesting property of Markov chain is that since it is a linear operator, if we increase (decrease) the rates of a node by $\alpha$ percent, based on the described procedure, after calculating the new steady state probabilities, the new value of that node will increase (decrease) by $\alpha$ percent. Further, since we assume a closed system, then we have a zero sum game. This means that a decrease (increase) in the $\pi_i$ will result to decrease or increase of $\pi_j$ for $j \neq i$ such that $\sum_{i=1}^{n} \pi_i = 1$.

It is important to mention that there is a pre-assumption in this manipulation of the original transition matrix that by slowing down the activity of a node, all of its connected industries redistribute their slack resources to other activities proportionally to the their flow rates. Therefore, here we assume that there is no limit in resources and production capacities or any limits on the absolute flow levels of money (commodity) over the edges of the network.

In the perturbation process, there might be nodes (assumingly with not a large structural power) that have effects on many other nodes. Thus instead of considering the total values of these effects, by focusing on the number of nodes that are being affected by the change in the activity of one node, we introduce the two following measures.

*Systemic Influence*, which is a measure for each economic node, calculated as the ratio of number of affected nodes (negatively or positively) to the total number of nodes, caused by a change in the activity of that node.

*Systemic Fragility*, which is a measure for each economic node, calculated as the ratio of number of times a node is affected (negatively or positively) by a change in the activity of all the other nodes.

Another possible sensitivity analysis is to consider the effect of each node on a global measure of the economic network such as Kemeny constant. This type of analysis sometimes leads to unexpected results, where by removing important nodes (in terms of steady state probabilities) the total flow of the network will improve and vice versa [10]. In the next section we will present the results of applying the above-mentioned analyses in to the evolving global economic network.

# Results

In this section, based on the previously described properties of the Markov chains we show the results of our experiments on the WIOD data set.

## The overall patterns of globalization

In this part, we focus on the global features of the underlying network by showing the results of mixing times and Kemeny constants for the years from 1995 to 2011. Fig 2 shows the sequence of mixing times for the corresponding Markov chains of each year. We run these iterations several times with the same threshold of termination for all the years and we observed that the mixing times are stable in different runs. This can be seen in the very small error bars around the average values of each year. As we expected the mixing time series has an overall downward pattern from 1995 to 2011, which indicates that during these years the underlying TMs and consequently the world economic network was getting more and more interconnected. Therefore, one could interpret this feature as an *index of globalization*. Further, as it is shown in Fig 2, this index reflects the effect of global financial crisis in 2008, which results to a jump in the mixing time in 2009. This implies that the world economic network was less connected in 2009 comparing to 2008. A similar pattern can be seen from 1997 to1999, where we could not argue its underlying reason. Nevertheless, the overall pattern shows a rapid globalization during 1995 to 2011, which seemingly will continue for the next coming years.

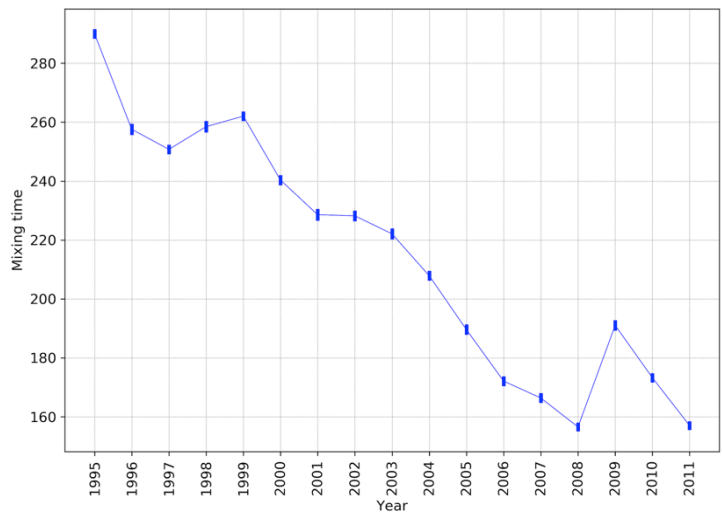

**Fig 2. The sequence of average mixing time of Markov chains as an aggregate** *index of globalization*

Lower values indicate more globally connected network. The error bars represent 3 standard deviations.

Similar to mixing times we expected that the Kemeny constant series to show a downward pattern. As presented in [5], we calculated the Kemeny constant of each Markov chain based on the Eigenvalue decomposition of the corresponding matrices. Fig 3 shows that although Kemeny constants have the same overall pattern as the time series of mixing times, including the shock in 2008 and 2009, there is an upward pattern in the values of Kemeny constants within the years 2000 to 2004. As a reminder, we should note that Kemeny constant indicates the average time from any given state (here any industry within any economy) to any random state in the network, where surprisingly this average time is constant independent of the starting point. However, when there is a local loop within the network this average time will increase. In the context of economic network, this might mean that within the years of 2000 to 2004, there might have been a creation or reinforcement of some local loops in the global economic network. Nevertheless, Kemeny constant is an aggregated and emergent measure of the underlying dynamics and one needs specific investigations in order to find out the underlying reasons for these macro behaviors.

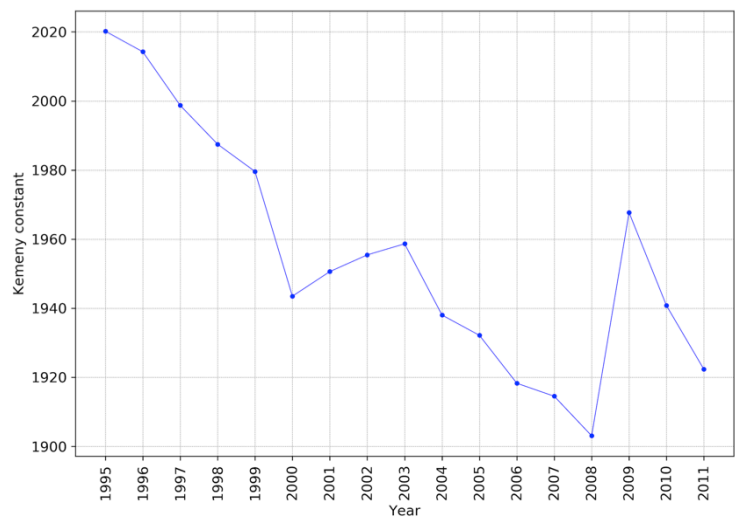

**Fig 3. The sequence of Kemeny constants of Markov chains as an aggregate** *index of globalization*

Lower values indicate more globally connected networks.

In the next section, we focus on the analysis of steady state probability distributions for different years.

## Steady state probabilities as a measure of structural power of economies

In a stochastic transition matrix, the first Eigenvector, $\pi$, shows the steady state probabilities of the underlying dynamical system. As described in the previous section, this can be calculated easily using power iteration method. However, as we discussed before, since we have one unique Markov chain for each year, we cannot claim that the underlying economic system reaches to its steady state within each year. However, the steady state vectors of each year can be interpreted as the *structural power* of each node in the economic network and since the structure of the network (i.e. the number of nodes in the global economic network) is fixed, comparing the time series of $\pi_i^y$ for each node $i$ at year $y$ reveals interesting results. Further, one can easily calculate different aggregated measures by summing up these steady state values over different categories such as industries or economies. As we will show there is a direct relation between the aggregated values of each economy, called $\pi_E^y$ and its GDP share at the same time. In principle, GDP as a measure of economy's welfare considers one economy in an isolated set up, while the steady state probabilities are being calculated based on the relationships between all the economic nodes. Therefore, looking at economies in isolation might reveal different results than considering the developments in other economies at the same time. Recently, in this direction there have been interesting works such as [11,15,16] that came up with measures of economic fitness of countries that are fundamentally relational and consequently reveal different features than classical GDP measures.

Fig 4 compares two time series of GDP shares of economies with the time series of $\pi_E^y$. Each individual plot corresponds to one aggregated economy (industries plus households/government), where the x-axis is the year and y-axis is for the GDP shares (red line) and $\pi_E$ values of each economy (blue lines). As we expected, the two time series are highly correlated. However, the differences between the two time series indicate an interesting aspect of these economies. We think this difference can be considered as a measure of *economic fitness* or the *structural potential* of the economies for further growth. As a hypothesis, we think whenever the GDP share is larger than the aggregated $\pi_E$ values of the economy, that economy is at risk (for example, the red gaps in Cyprus and Greece) and when the gap is blue, this means that the country has still more potential structural power than what is being produced. An interesting feature of this ratio is that it does not correlate with the overall patterns of the economy in time. For example, while Germany and Japan are loosing their global competitiveness (with downward patterns), still they are not in a risky area (the blue gap). On the other hand, while India and Turkey for example are gaining more competitive powers (with upward patterns), both are at the same time going to the risky area (the red gap).

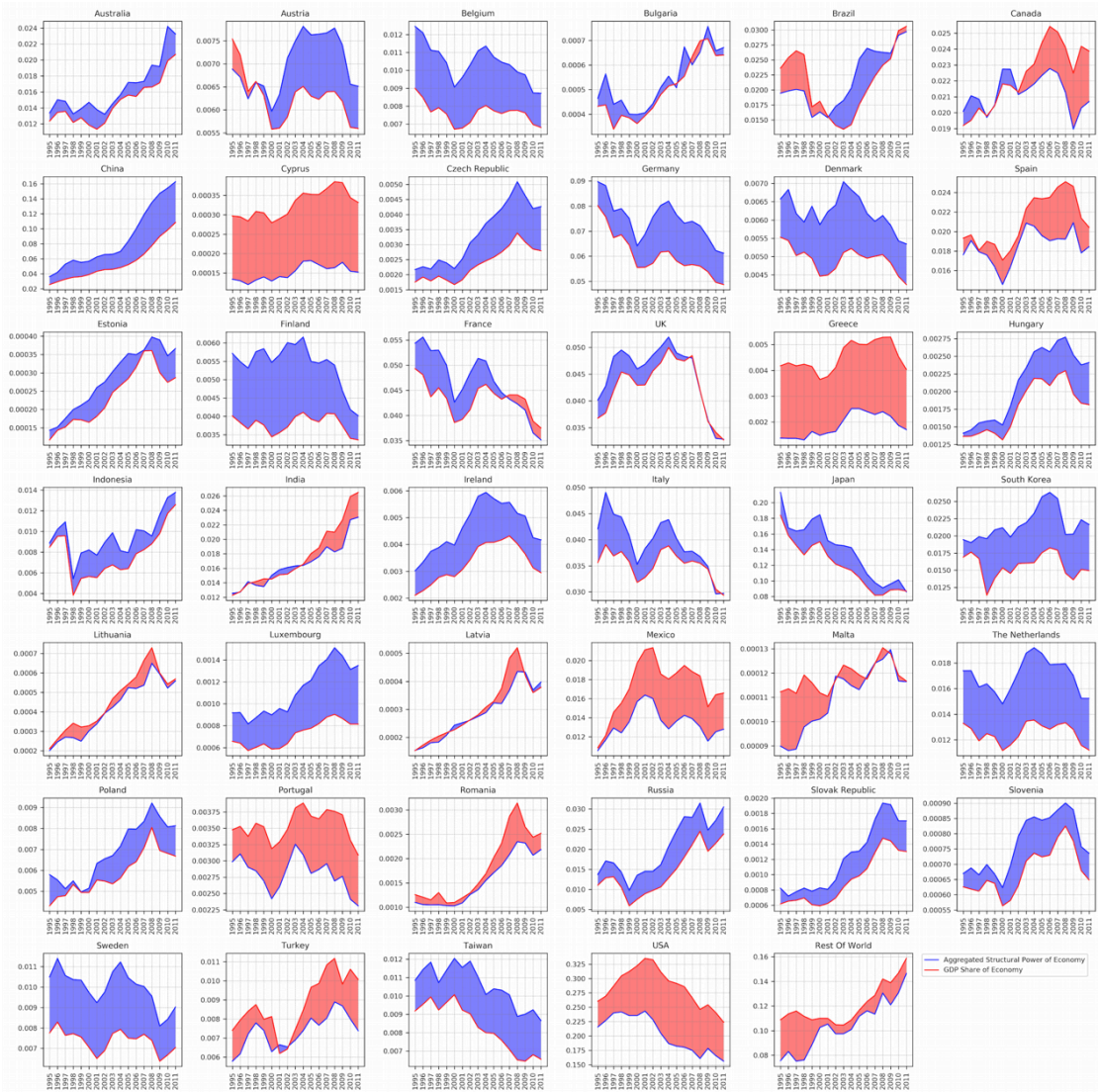

**Fig 4. GDP shares of economies (red line) compared with their aggregated structural powers (blue line) over time**

The ratio between two time series reveals the *structural potential* of the economies for further growth (blue gaps) or the risk of economic failure (red gaps).

Within the literature of economic complexity there has been always an interest in predicting the future states of the dynamical systems. Plotting the patterns of the so-called BRIC countries (Brazil, Russia, India and China) together shows an interesting similarity (Fig 5). It seems that all of these countries have passed a curve shape behavior and in 2011 they are slowing down, where the GDP share is getting closer to the aggregated $\pi_E$, hypothetically implying less structural potential for further growth. The red dashed lines are calculated based on the moving average of the first momentum of each time series with the time lags between 3 to 6 years. The ticker dashed line shows the median prediction.

This result is similar to the results of the recent works published in [16], where the authors predict the future economic fitness of different economies in comparison to their GDP per capita.

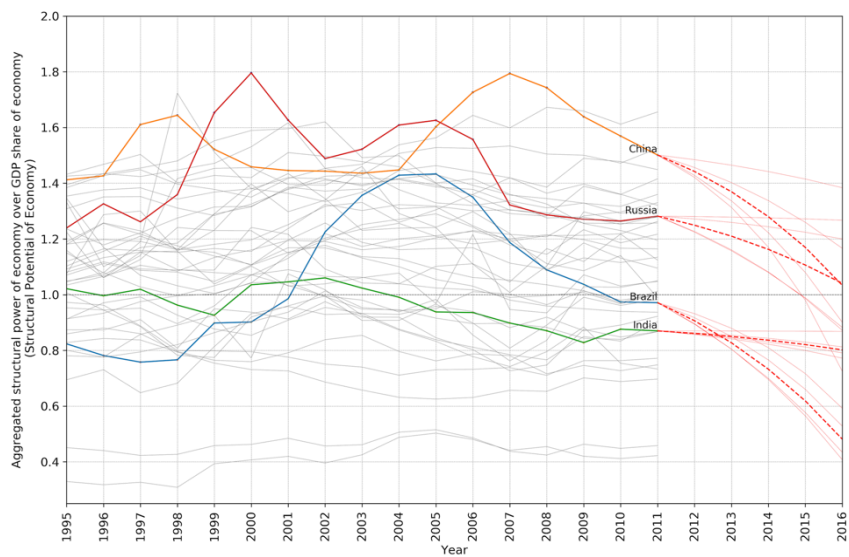

**Fig 5. Predicted trends of structural potential of different economies**

In the next section, we will focus on the sensitivity analysis of the Markov chains in order to assess the influence of different nodes on each other and to further, identify those fragile nodes that get affected by shocks in the network. In addition, we analyze the effect of slowdown in the economic activities of each individual node on the overall monetary flow of the world economic network.

# Sensitivity analysis of Markov chains

Since Markov chain provides a formalism of the underlying dynamical system, as described before, it is then very easy to perform sensitivity analysis by slight changes in the values of the constructed transition matrix.

In a drastic scenario, Fig 6 shows the effect of 99% slow down in the electrical and optical equipment industry of China in 1995 and 2011 respectively. As it was expected, comparing to 1995, a change in this industry in 2011 has enormous negative and positive effects on the final shares (based on the new $\pi_t$ vector) of other industries across the globe. In the depicted diagrams, negative effects are highlighted by red color and positive effects are shown by green color. The size of green or red circles is proportional to the primary values in $\pi_t$ vector of that economic node. For better visualization purpose, those nodes with less than 1 percent of change in their corresponding structural power ($\pi_{t,i}$) are shown with a small dot. A large orange circle highlights the perturbed industry. Further, it is important to mention that the nodes are arranged in a two dimensional space, based on their similarities in exports related links. This means, closer nodes have similar export patterns.

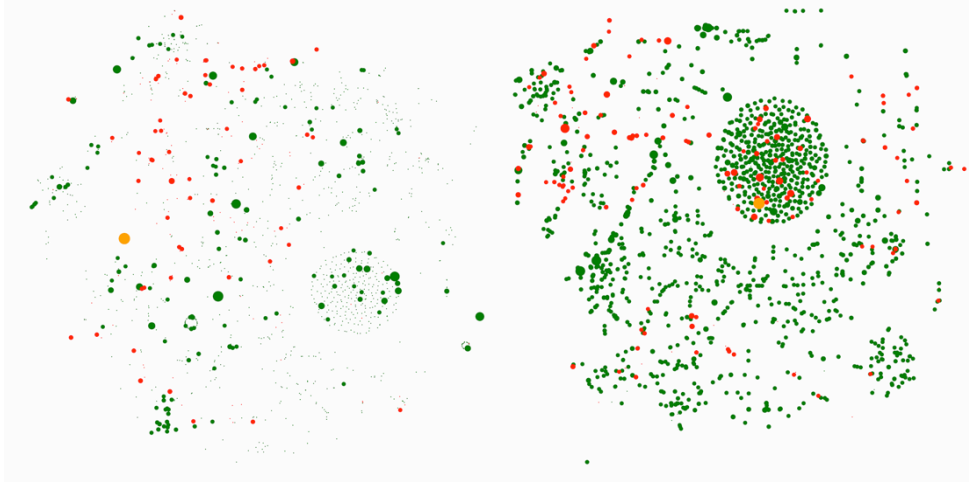

**Fig 6. The effect of 99% slowdown of electrical and optical equipment industry of China**

Left side shows the shocked network in 1995 the right side shows 2011. The green (red) color declares an increase (decrease) in the final share (structural power, $\pi_{t,i}$) of the node as a result of the slow down in the selected industry.

As we mentioned before, by changing $\alpha$ percent of the activity of node $i$, the total absolute amount of positive and negative changes (i.e. redistribution of values in new $\pi_t$) are equal to $\alpha$ percent of $\pi_{t,i}$. Therefore, the global effect of fluctuations in the activity of each node is the same as the change in its $\pi_i$.

Thus instead of focusing on the magnitude of changes, two new measures (*Systemic Influence* and *Systemic Fragility*) that were introduced in the previous section are based on the multitude of changes, happening as a result of a shock in the network.

In order to calculate these two measures for all the nodes in the network, we performed the perturbation procedure, which we described before for all the nodes with $\alpha = -99$. Fig 7 shows the distribution of systemic fragility vs. systemic influence of each node as a result of 99% percent slowdown of each individual node for the year 2011. We should note that for the calculation of these two measures we only considered those absolute changes, which are more than 0.5% of the structural power of the node itself. The size and color of nodes correspond to their structural powers (i.e. $\pi_i$ or Eigen Centralities). As it can be seen although all the nodes with high structural power have relatively high systemic influence, there are nodes with high systemic influence, but low structural power. For the case of systemic fragility there is even less correlations to structural power. While nodes with high structural power are relatively robust (i.e. low fragility), there is a very wide range of fragility values for those nodes with low structural power.

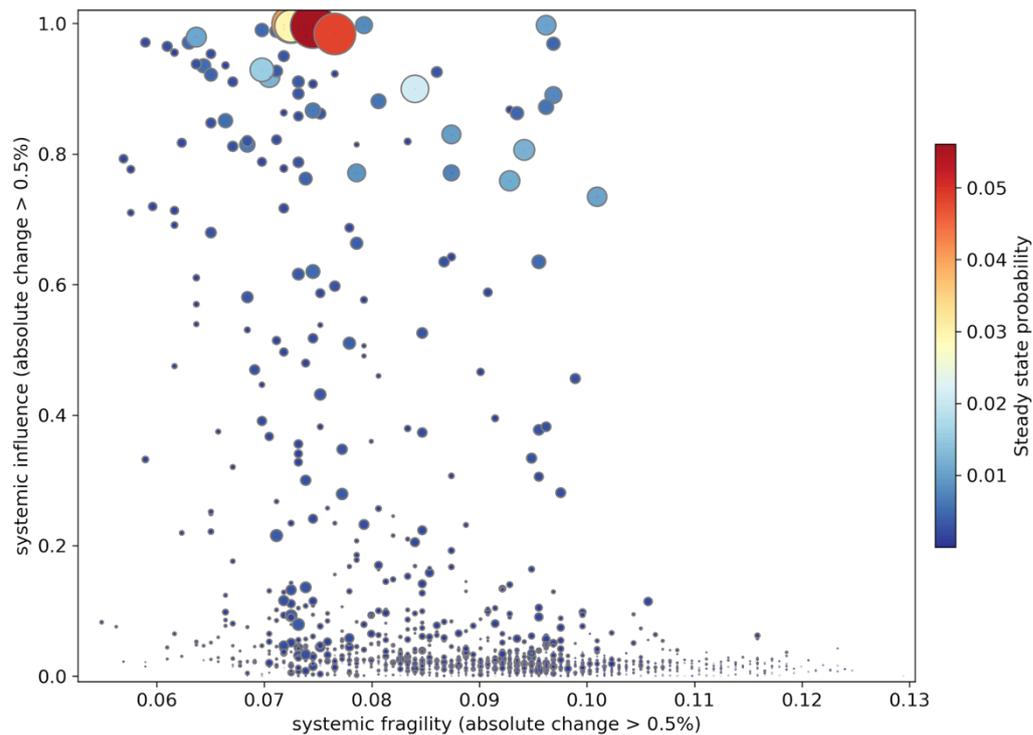

**Fig 7. Systemic fragility vs. systemic influence of each industry for the year 2011**

Table 1 shows the top 10 economic nodes (except Rest Of World) in 2011 with the highest systemic influences along with their structural power and systemic fragility.

**Table 1. Top 10 nodes with the highest systemic influence in 2011**

| Rank | Names | Structural power | Systemic Fragility | Systemic Influence |
|------|-------|------------------|--------------------|--------------------|
| 1 | China-Government | 0.0407637 | 0.0724932 | **0.998645** |
| 2 | Japan-Real Estate Activities | 0.00434353 | 0.0718157 | **0.997967** |
| 3 | Brazil-Government | 0.0106923 | 0.096206 | **0.99729** |
| 4 | India-Government | 0.00788957 | 0.0792683 | **0.99729** |
| 5 | USA-Government | 0.0560428 | 0.0745257 | **0.995935** |
| 6 | USA-Real Estate Activities | 0.00796444 | 0.0738482 | **0.995935** |
| 7 | Japan-Government | 0.0294709 | 0.0724932 | **0.995935** |
| 8 | USA-Retail Trade, Except of Motor Vehicles | 0.0049287 | 0.0731707 | **0.995257** |
| 9 | USA-Wholesale Trade and Commission Trade | 0.00488901 | 0.0738482 | **0.99187** |
| 10 | USA-Renting of M&Eq and Other Business Activities | 0.0116091 | 0.0718157 | **0.99187** |

Further, Table 2 shows the top 10 economic nodes in 2011 with the lowest systemic fragilities. Note that since we are interested in the nodes that can be attributed to a specific economy we removed those nodes, which were attributed to Rest of the World (ROW).

**Table 2. Top 10 nodes (except rest of the world) with the lowest systemic fragility in 2011**

| Rank | Names | Structural power | Systemic Fragility | Systemic Influence |
|------|-------|------------------|--------------------|--------------------|
| 1 | UK-Electrical and Optical Equipment | 0.000292468 | **0.054878** | 0.0826558 |
| 2 | Finland-Electrical and Optical Equipment | 0.000137664 | **0.0562331** | 0.0758808 |
| 3 | Taiwan-Manufacturing; Recycling | 3.36E-05 | **0.0569106** | 0.0223577 |
| 4 | Germany-Electrical and Optical Equipment | 0.00154004 | **0.0575881** | 0.776423 |
| 5 | Germany-Chemicals and Chemical Products | 0.00110951 | **0.0575881** | 0.710027 |
| 6 | Ireland-Machinery | 1.32E-05 | **0.0589431** | 0.0149051 |
| 7 | USA-Electrical and Optical Equipment | 0.00232597 | **0.0589431** | 0.970867 |
| 8 | Malta-Electrical and Optical Equipment | 6.46E-06 | **0.0589431** | 0.0216802 |
| 9 | Germany-Machinery | 0.00183113 | **0.0596206** | 0.719512 |
| 10 | Denmark-Chemicals and Chemical Products | 7.48E-05 | **0.0609756** | 0.0264228 |

As another possible sensitivity analysis, we assessed the role of each individual economic node to the overall flow of the economic network. As we discussed before, Kemeny constant and mixing time are two global measures of a Markov chain, where the lower values show a more globally connected network and faster flow of money. In [25], the authors introduce a simple procedure to see the effect of removing each node on the average flow within a network. In the domain of urban traffic network analysis, this method has been used to analyze the effect of closing a road (or a junction) on the overall flow of the network, where the results are sometimes paradoxical. In [25, 26] it has been shown that by removing some nodes with high structural power (i.e. high level of expected share of traffic) the overall average flow (in terms of Kemeny constant) will be better. This phenomenon is known as Braess paradox [27]. This apparently paradoxical result implies that in order to improve the overall flow of a network, some times it is better not to add a new node, but to remove some.

We implemented this procedure to the economic network for all the years from 1995 to 2011, where we calculated the percent of change in the Kemeny constant of the Markov chain by slowing down the activity of each node by 99%. We manipulated the transition matrix with the same procedure that we used for the calculation of systemic influence and systemic fragility. For the year 2011 Fig 8 shows the relationship between steady state probabilities (i.e. structural power of economic nodes) and the percent of change in Kemeny constant, caused by the slowdown in the activity of the node. As we expected, the slowdown of the majority of economic nodes, leads to higher Kemeny constants (i.e. a slower overall monetary flow across the network). On the other hand, there are some nodes that the slow downs of their activity decrease the Kemeny constant (i.e. a faster overall monetary flow across the network).

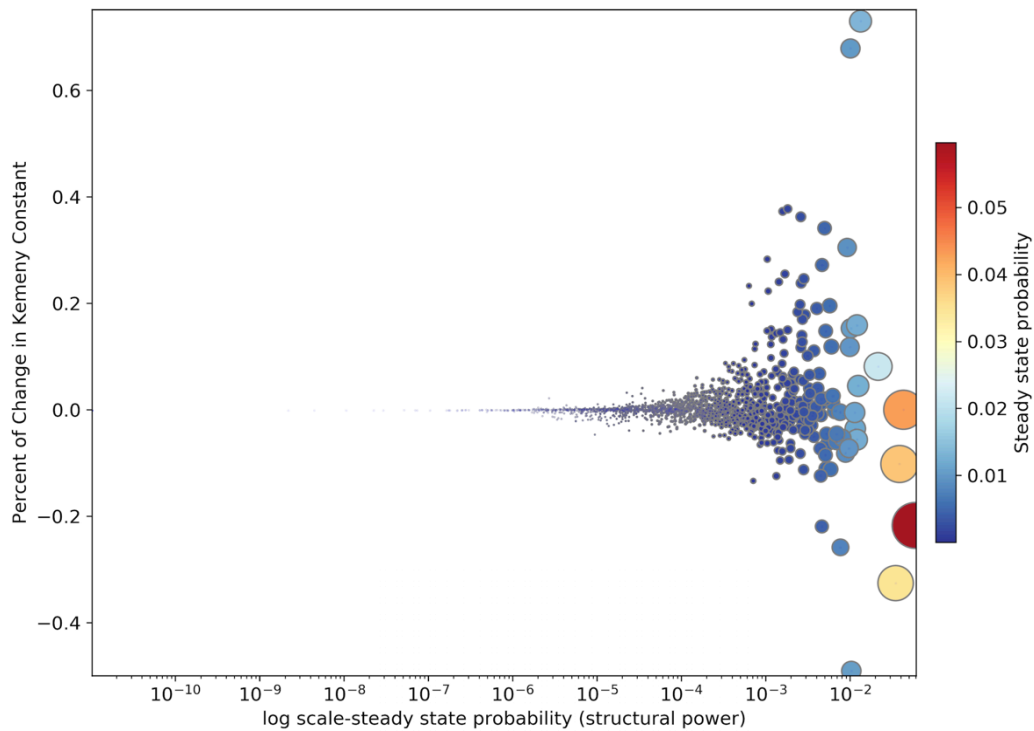

**Fig 8. Effect of slowing downs the activity of economic nodes on Kemeny constant in 2011**

In addition, Table 3 shows top 10 economic nodes with the highest positive effects on Kemeny constant in 2011 along with their corresponding structural power (and their systemic influences as calculated before. Similar to previous tables, we removed those nodes related to Rest of World in the following tables.

**Table 3. Top 10 nodes (except rest of the world) with the highest positive effects on Kemeny constant in 2011**

| Rank | Names | Structural power | Systemic Influence | % of change in Kemeny constant |
|---|---|---|---|---|
| 1 | China-Electrical and Optical Equipment | 0.0109195 | 0.978997 | **0.637486** |
| 2 | Germany-Transport Equipment | 0.00282169 | 0.811653 | **0.354232** |

| Rank | Names | Structural power | Systemic Influence | % of change in Kemeny constant |
|---|---|---|---|---|
| 3 | China-Textiles and Textile Products | 0.00498553 | 0.970867 | **0.2612** |
| 4 | Germany-Machinery | 0.00183113 | 0.719512 | **0.255455** |
| 5 | Germany-Electrical and Optical Equipment | 0.00154004 | 0.776423 | **0.230203** |
| 6 | Germany-Chemicals and Chemical Products | 0.00110951 | 0.710027 | **0.209375** |
| 7 | Romania-Government | 0.000723924 | 0.0765583 | **0.203782** |
| 8 | Russia-Government | 0.0104454 | 0.734417 | **0.202094** |
| 9 | USA-Transport Equipment | 0.00252037 | 0.910569 | **0.191263** |
| 10 | USA-Coke, Refined Petroleum and Nuclear | 0.00264798 | 0.469512 | **0.182051** |

Further, Table 4 shows the top 10 economic nodes with the highest negative effects on Kemeny constant in 2011 along with their corresponding Eigen centralities and systemic influences.

**Table 4- Top 10 nodes (except rest of the world) with the highest negative effects on Kemeny constant in 2011**

| Rank | Names | Structural power | Systemic Influence | % of change in Kemeny constant |
|---|---|---|---|---|
| 1 | Brazil-Government | 0.0106923 | 0.99729 | **-0.448618** |
| 2 | Japan-Government | 0.0294709 | 0.995935 | **-0.28818** |
| 3 | India-Government | 0.00788957 | 0.99729 | **-0.275446** |
| 4 | Mexico-Government | 0.00473883 | 0.968835 | **-0.201355** |
| 5 | USA-Government | 0.0560428 | 0.995935 | **-0.201247** |
| 6 | Greece-Government | 0.000652282 | 0.0623306 | **-0.133354** |
| 7 | Finland-Government | 0.00127287 | 0.105014 | **-0.116016** |
| 8 | Spain-Government | 0.00610205 | 0.871951 | **-0.114455** |
| 9 | China-Government | 0.0407637 | 0.998645 | **-0.111065** |
| 10 | Sweden-Government | 0.0030026 | 0.377371 | **-0.110952** |

Unlike Table 3, in Table 4 all the top nodes are related to governments, where by slowing down their activities, one can expect to have a better overall flow in the network (i.e. smaller Kemeny constant). In order to investigate if there are other nodes from different sectors than government that will show this paradoxical effect, Table 5 shows the top 10 nodes without governments and rest of the world (ROW), whose economic slowdowns improve the overall flow of money (i.e. smaller Kemeny constant). While, comparing to values in Table 4, the percentages of changes in Table 5 are much smaller, it is interesting to note that five out of 10 top nodes belong to the sector of real estate activities.

**Table 5 - Top 10 nodes (except governments and rest of the world) with the highest negative effects on Kemeny constant in 2011**

| Rank | Names | Structural power | Systemic Influence | % of change in Kemeny constant |
|---|---|---|---|---|
| 1 | Japan-**Real Estate Activities** | 0.00434353 | 0.997967 | **-0.0994393** |
| 2 | Brazil-Public Admin and Defense; Compulsory So... | 0.00153962 | 0.0264228 | **-0.0872356** |
| 3 | India-Agriculture, Hunting, Forestry and Fishing | 0.00157759 | 0.0528455 | **-0.0769576** |
| 4 | USA-**Real Estate Activities** | 0.00796444 | 0.995935 | **-0.0739146** |
| 5 | Australia-**Real Estate Activities** | 0.00132178 | 0.0121951 | **-0.0690959** |
| 6 | France-**Real Estate Activities** | 0.00190527 | 0.00880759 | **-0.0678286** |
| 7 | Japan-Public Admin and Defense | 0.00371188 | 0.910569 | **-0.0646992** |
| 8 | Brazil-**Real Estate Activities** | 0.000888012 | 0.0216802 | **-0.0593976** |
| 9 | Japan-Renting of M&Eq and Other Business Activities | 0.00458614 | 0.988482 | **-0.0576155** |
| 10 | Sweden-Government | 0.0030026 | 0.377371 | **-0.110952** |

The results shown here are based on world economic network in 2011. However, in the same way as the previous measures, it is possible to analyze the behavior of these measures over time that we leave it to future research. Fig 9 shows that the slowdown of the activity of economic nodes has most of the times a very little positive effect on the Kemeny constant of the corresponding year. On the other hand there are few nodes whose changes have a big impact (either negative or positive) on the overall flow of the money in the global economic network during the years from 1995 to 2011. It is also interesting to see how the influence of China's electrical an optical equipment industry has increased during the last decade, which is presumably because of expansion of information technology across the world.

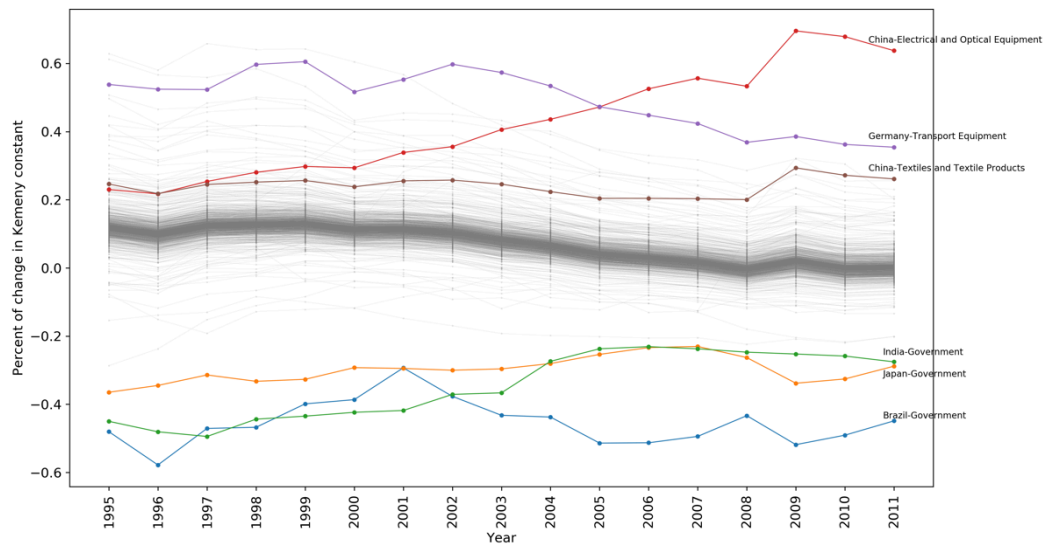

**Fig 9. The paradoxical effect of slowdown in the activity economic nodes (except rest of the world) on the Kemeny constants**

## Discussions and Conclusions

Thanks to the recently available World Input Output Database (WIOD), in this work we modeled the evolution of world economic network from 1995 to 2011 by a series of finite state Markov chains. As a result, we were able to analyze different aspects of the underlying dynamical system, by analyzing different properties of the constructed Markov chains.

We showed that the ratio between the aggregated steady state probabilities of an economy to its GDP-share could be considered as a measure of structural potential of economies for further growth, where the values less than one show a decline in the speed of growth (economic slowdown) and the values more than one show the potential for faster economic growth. Further, we claimed that this ratio could be considered as a risk measure, which is independent of the trend an economy has in comparison to other economies. Therefore, there are economies gaining more structural power with a risky path (i.e. lower structural power than the GDP share) and vice versa.

In addition, via perturbation analysis of the underlying transition matrices we introduced two measures of systemic risk, called *systemic influence* and *systemic fragility,* which measure the effect of change in the activity of one node (i.e. an industrial sector of an economy) on the structural power of all the other nodes in terms of multitude rather than the magnitude. Further, we showed that the slow down of activities in different nodes has both negative and positive results in terms of Kemeny constant, which is a measure of connectivity of the network. This result, which is paradoxical, needs further investigations.

Finally, we should mention that there are similarities between our work and two recent works [15,16], where using a bi-partite economy-product network, they come up with an interesting measure of economic fitness and product complexity. However, we think our approach based on the Markov chain formalism on input-output tables has more advantages. We will investigate these relations in our future research.

# Supporting information

All the data sets and codes used to produce the results of this work can be found at

https://sevamoo.github.io/Markovian_IO_SI_PLOSONE/